\documentclass[twocolumn]{aastex62}

\graphicspath{{./}{figures/}}

\received{XXX}
\revised{YYY}
\accepted{ZZZ}
\submitjournal{AJ}

\usepackage{graphicx}	
\usepackage{amssymb}	
\usepackage[inline]{enumitem}
\usepackage{gensymb}

\shorttitle{Destroyers of Worlds}
\shortauthors{Stephan, Naoz, and Gaudi}

\begin{document}

\title{A-type Stars, the Destroyers of Worlds:  The lives and deaths of Jupiters in evolving stellar binaries}

\correspondingauthor{Alexander P. Stephan}
\email{alexpstephan@astro.ucla.edu}

\author[0000-0001-8220-0548]{Alexander P. Stephan}
\affil{Department of Physics and Astronomy, University of California, Los Angeles, Los Angeles, CA 90095, USA}
\affiliation{Mani L. Bhaumik Institute for Theoretical Physics, University of California, Los Angeles, Los Angeles, CA 90095, USA}

\author[0000-0002-9802-9279]{Smadar Naoz}
\affiliation{Department of Physics and Astronomy, University of California, Los Angeles, Los Angeles, CA 90095, USA}
\affiliation{Mani L. Bhaumik Institute for Theoretical Physics, University of California, Los Angeles, Los Angeles, CA 90095, USA}

\author[0000-0003-0395-9869]{B. Scott Gaudi}
\affiliation{Department of Astronomy, The Ohio State University, Columbus, OH 43210, USA}



\begin{abstract}

Hot Jupiters (HJs), gas giant planets orbiting their host stars with periods on the order of days, commonly occur in the Galaxy, including relatively massive ($1.6-2.4$~$M_\odot$, i.e., A-type main sequence stars) and evolved stars. The majority of A-type main sequence stars have stellar binary companions, which can strongly affect the dynamical evolution of planets around either star. In this work, we investigate the effects of gravitational perturbations by a far away stellar companion on the orbital evolution of gas giant planets orbiting A-type stars, the so-called Eccentric Kozai-Lidov (EKL) mechanism, including the effects of general relativity, post-main sequence stellar evolution, and tides. We find that only $0.15~\%$ of A-type stars will host HJs during their main sequence lifetime. However, we also find a new class of planets, Temporary Hot Jupiters (THJs), that form during the post-main sequence lifetime of about $3.7~\%$ of former A-type main sequence stars. These THJs orbit on periods of tens to a hundred days and only exist for a few $100{,}000$~years before they are engulfed, but they reach similar temperatures as ``classical'' HJs due to the increased stellar luminosities. THJs' spin-orbit angles will mostly be misaligned. {THJ effects on the host stars' evolution could also be observable for longer than a few $100{,}000$~years.}
Overall, we find that approximately $70~\%$ of all gas giant planets orbiting A-type stars will eventually be destroyed or engulfed by their star, about $25~\%$ during the main sequence lifetime, about $45~\%$ during post-main sequence evolution.

\end{abstract}



\section{Introduction}\label{sec:Intro}

Exoplanets have been observed around a variety of host stars, with different masses, at all stages of stellar evolution, including main-sequence, sub giant and red giant branch (RGB) stages \citep[e.g.,][]{Charpinet+2011,Johnson+2011A,Howard+2012,Gettel+2012,Barnes+2013,Nowak+2013,Reffert+2015,Niedzielski+2015,Niedzielski+2016}.
However, for {massive, evolved} stars, there appears to be a deficit in short-period or high eccentricity planets \citep[e.g.,][]{Sato+2008,Sato+2013,Bowler+2010,Johnson+2007,Johnson+2008,Johnson+2010A,Johnson+2010B,SchlaufmanWinn2013}. Furthermore, high metal abundances in so-called ``polluted'' white dwarf (WD) atmospheres indicate the presence of the remnants of planetary systems around these stars; the processes by which this material was brought onto the WDs is an active topic of research \citep[e.g.,][]{Farihi+2009,Farihi+2010,Klein+2010,Klein+2011,Melis+2011,Zuckerman+2011,Xu+2013,Xu+2017,Stephan+2017}.

The architectures of these planetary systems have become the focus of a rich field of research, as it was recently shown that dynamical processes play an important role in planetary system formation and evolution. These processes include resonant interactions \citep[e.g.,][]{LithwickWu2012,BatyginMorbidelli2013,Petrovich+2013,GoldreichSchlichting2014}, planet-planet scattering \citep[e.g.,][]{RasioFord1996,Nagasawa+2008,Chatterjee+2008,Boley+12,BN2012} and secular perturbations from a companion \citep[either a star or a planet; e.g.,][]{Fabrycky+07,Wu+07,Naoz11,Naoz+12bin,Naoz+11sec} or from multiple planets \citep[e.g.,][]{WuLithwick2011,Denham+2018}.

A particularly interesting group of discovered exoplanets are the so called ``Hot Jupiters'' (HJs), which are gas giants that orbit their host stars on very tight orbits with periods on the order of a few days. While this class of exoplanets seems ubiquitous in the galaxy, it is noticeably absent in our own solar system. Several models have been developed to explain the formation and existence of these HJs, including gravitational perturbations of the planets' original orbits to high eccentricities, followed by tidal dissipation and orbit circularization and shrinking \citep[e.g.,][]{Dan,Naoz11,Naoz+12bin,BN2012,Petrovich2015, Frewen+2016}, as well as disk migration during giant planet formation \citep[e.g.,][]{Armitage+2002, Mass+03, Armitage2007}. The idea that outer companions have perturbed these planets and led to high eccentricity migration is also supported by recent observational campaigns that have shown that most HJs have a far away companion, either a star or a planet \citep[e.g.,][]{Knutson+2014}, though it remains unclear if most of these companions can trigger high-eccentricity migration \citep[e.g.,][]{Ngo+2016}. For a three body system consisting of star-planet-star to be long-term stable the inner two bodies, the main star and the gas giant, have to be on a much tighter orbit than the third, outer, object, leading to a hierarchical configuration.

In recent years, so-called ``retired'' A-type stars, {which are observed as K-type giants,} have been focused by many studies in attempts to discover exoplanets \citep[e.g.,][]{Johnson+2007,Johnson+2008,Bowler+2010,Johnson+2010B,Johnson+2011A,Johnson+2011B}. ``Retired'' A-type stars are stars that would be classified as A-type during their main-sequence lifetime, but which have evolved beyond the main sequence and are sub-giant or giant {K-type} stars at their currently observed life stage. While main-sequence A-type stars usually rotate rapidly and have high surface temperatures, greatly impeding exoplanet detection through radial velocity measurements, ``retired'' A-type stars rotate slower, are cooler, and have allowed for the discovery of several exoplanets. Furthermore, several HJs have also been discovered around A-type main sequence stars through exoplanet transits \citep[e.g.,][]{Gaudi+2017,MarshallJohnson+2018}. 

The classification of ``retired'' A-type stars has been shown to be rather challenging, as different methods to determine stellar masses seem to yield different values \citep[e.g.,][]{Lloyd2011,Johnson+2013,North+2017,Stello+2017,Ghezzi+2018}{, implying that some of these stars might rather be ``retired'' F-type stars}. There seems to be uncertainty over the validity of some methods when compared to precise astroseismological measurements. In this work we avoid these classification problems by simply focusing on a particular range of stellar masses, between $1.6$ and $3$~$M_\odot$ (with masses above $\sim2.4$~$M_\odot$ technically belonging to the low-end B-type mass range). This intermediate stellar mass range broadly coincides with classical definitions of main-sequence A-type star masses \citep{Adelman2004} and we refer to this mass range when labelling a star as A-type. A further discussion concerning A-type star evolution is given in Appendix \ref{App:AppendixA}. 

Stars more massive than the sun, like A-type stars, reach post-main sequence evolution much faster than smaller stars, and the vast majority of them has stellar companions \citep[e.g.,][]{Raghavan+10,Moe+2017,Murphy+2018}. Indeed, several A-type stars with HJs have companion stars \citep[e.g.,][]{MarshallJohnson+2018,Siverd+2018}. This leads to an interesting interplay of dynamical and stellar evolution effects that needs to be considered for planets in such systems. On the one hand, a hierarchical star-planet-star configuration will lead to secular oscillations of the orbital parameters due to gravitational perturbations by the outer companion on the planet's orbit, often leading to extreme eccentricities, the so-called Eccentric Kozai-Lidov (EKL) mechanism \citep{Kozai,Lidov,Naoz2016}. On the other hand, post-main sequence stellar evolution will lead to, for example, stronger tidal dissipation or engulfment of close-in planets due to stellar radial expansion, and expanded orbits due to stellar mass loss. In this work, we study the combined interplay of these dynamical and stellar effects for Jupiter-sized planets in stellar binaries with A-type star primaries. {We find that short stellar evolution timescales, high prevalence of binary companions, and strong tides during post-main sequence evolution result in the destruction of nearly $70~\%$ of Jupiters orbiting A-type stars, more than we would expect for lower mass single stars.} We propose that observed planets around intermediate mass main and post-main sequence stars with stellar companions\footnote{Planetary companions might have different results.} are consistent with our predicted results.

\section{Numerical Setup}\label{sec:Num}

We perform large Monte-Carlo simulations that follow the dynamical evolution of hierarchical three-body systems, consisting of a relatively tight {\it inner binary} pair of a star and a Jupiter-sized planet, which are orbited by another star on a distant orbit as {\it outer binary}. We solve the hierarchical secular triple equations up to the octupole level of approximation \citep[the so-called EKL mechanism, e.g.,][]{Naoz2016}, including general relativity effects on both inner and outer orbits \citep[e.g.,][]{Naoz+12GR}, static tides between the primary star and the planet (following \citealt{Hut} and \citealt{1998EKH}; see \citealt{Naoz2016} for the complete set of equations). {Tides for radiative stars are also estimated to be much weaker than for convective stars, so we use different tidal models for (radiative) main sequence and (convective) red giant stars \citep[e.g.,][]{Zahn1977}; however observations are uncertain about their distinctiveness \citep{CollierCameron+2018}}. {The tidal Love numbers for stars and gas giants are set to 0.014 and 0.25, respectively \citep{1998KEM}, and we choose a viscous timescale of $1.5$~years for both.} {We also include} the effects of stellar evolution on masses, radii, and spins on the two stars, as derived from the stellar evolution code {\tt SSE} by \citet{Hurley+00}.
Unlike G and F-type main sequence stars, which exhibit magnetic braking due to their convective envelopes, A-type main sequence stars are nearly completely radiative and do not experience significant magnetic braking \citep{vanSaders+2013}. Their spin rates therefore do not substantially slow down during their main sequence evolution lifetime{, however, magnetic braking can occur during the red giant phase after a convective envelope has formed, additionally to the slowing of the spin due to stellar expansion. These factors are included in the calculations performed with {\tt SSE} and we also switch between tidal models for radiative and convective stars based on {\tt SSE} determinations of evolutionary phases. Overall, the main differences between A-type stars and smaller stars lie in the much more rapid stellar rotation rate and the weaker tidal dissipation for radiative stars, which weaken the importance of stellar tides during the main sequence lifetime. During post-main sequence evolution however, the massively expanded stellar radii, together with the more convective nature of red giants, greatly increase the strength of stellar tidal dissipation, beyond the strength of tides for less massive post-main sequence stars.}
The interplay between EKL mechanism and stellar evolution has previously been shown to play an important role in shaping the underlying dynamics and outcome of these systems \citep[e.g.,][]{Kratter+12,Michaely+14,Shappee+13,Frewen+2016,Naoz+2016,Stephan+2016,Stephan+2017,Toonen+2016}.

The mass of the primary star, $m_{\star,1}$, is taken from a Salpeter distribution with $\alpha=2.35$ \citep{Salpeter1955}, however the mass range is restricted between $1.6$ and $3.0$~$M_\odot$ in order to ensure that the planet host star is an A-type or, at most, a small B-type star during its main sequence lifetime \citep{Adelman2004}. 
The stellar initial radii and spins are calculated using {\tt SSE} by \citet{Hurley+00}. Each primary star is given one planet, whose mass ($m_p$), size, and spin are set equal to those of Jupiter. The mass of the outer companion star, $m_{\star,2}$, is determined by the binary mass ratio distribution taken from \citet{Duquennoy+91}. The semi-major axis, $a_1$ of the {\it inner binary} of A-type star and Jupiter-sized planet is chosen uniformly between $1$ and $10$~AU, while the {\it outer binary} orbit's semi-major axis, $a_2$, is again taken from the distribution in \citet{Duquennoy+91} for the stellar companion. The inner orbit's eccentricity is initially set to a small value, $e_1=0.01$, as are the stellar and planetary spin orbit angles, since we assume that the planet was formed in a gaseous disk, while the outer orbit's eccentricity, $e_2$ is chosen uniformly between $0$ and $1$. To ensure long-term stability we reject systems where $a_2$ is greater than $\sim 10{,}000$~AU, as galactic tides will tend to separate such systems relatively quickly \citep{Kaib+2013}, and we only consider hierarchical systems to ensure long-term stability, which requires: 
\begin{equation}
    \epsilon=\frac{a_{1}}{a_2}\frac{e_2}{1-e_2^2}<0.1 \ ,
\label{eq:stability}
\end{equation}
and ${a_{1}}/{a_2}<0.1$ \citep[e.g.,][]{Naoz2016}. The inclination, $i$ between the inner and outer orbit's angular momenta is chosen isotropically in cosine. 

In total, we simulate $4{,}070$ systems using these parameter conditions, $3{,}000$ of which have a primary star mass smaller than about $2.4$~$M_\odot$ and therefore safely classify as A-type stars during their main sequence lifetime. We calculate the dynamical evolution of these systems for $13$~Gyr, or until a stopping condition is fulfilled. If a planet touches the surface of a star or crosses the Roche limit, we stop the integration. We define the Roche limit $R_{Roche,A}$ of a body of mass $m_A$ and radius $r_A$ in respect to an orbiting body of mass $m_B$ as: 
\begin{equation}
    R_{Roche,A}=1.66 \times {r_A} \left( \frac{m_A+m_B}{m_A} \right)^{1/3}
\label{eq:Roche}
\end{equation}
Note that this equation simplifies to $R_{Roche,A}=1.66 \times {r_A}$ and $R_{Roche,B}=1.66 \times {r_B}\left( \frac{m_A}{m_B} \right)^{1/3}$ if $m_A >> m_B$, as in the case of a star with mass $m_A$ being orbited by a much smaller planet of mass $m_B$. The disruption of planets is highly sensitive to the Roche limit and thus to the numerical pre-factor value, chosen here to be $1.66$. Numerical simulations by  \citet{Guillochon+2011} and \citet{Liu+2013} suggested a larger value, (i.e., $2.7$) while \citet{Faber+2005} simulations found $\sim 2.2$. Thus, choosing here a fiducial value of $1.66$ means that the number of engulfed planets represents a lower limit on the fraction of planets that can be engulfed during post-main sequence radial stellar expansion. On the other hand, this value might lead us to overestimate the number of planets forming HJs during the main sequence lifetime \citep[see][]{Petrovich2015}.

\begin{figure*}
\hspace{0.0\linewidth}
\includegraphics[width=1.\linewidth ]{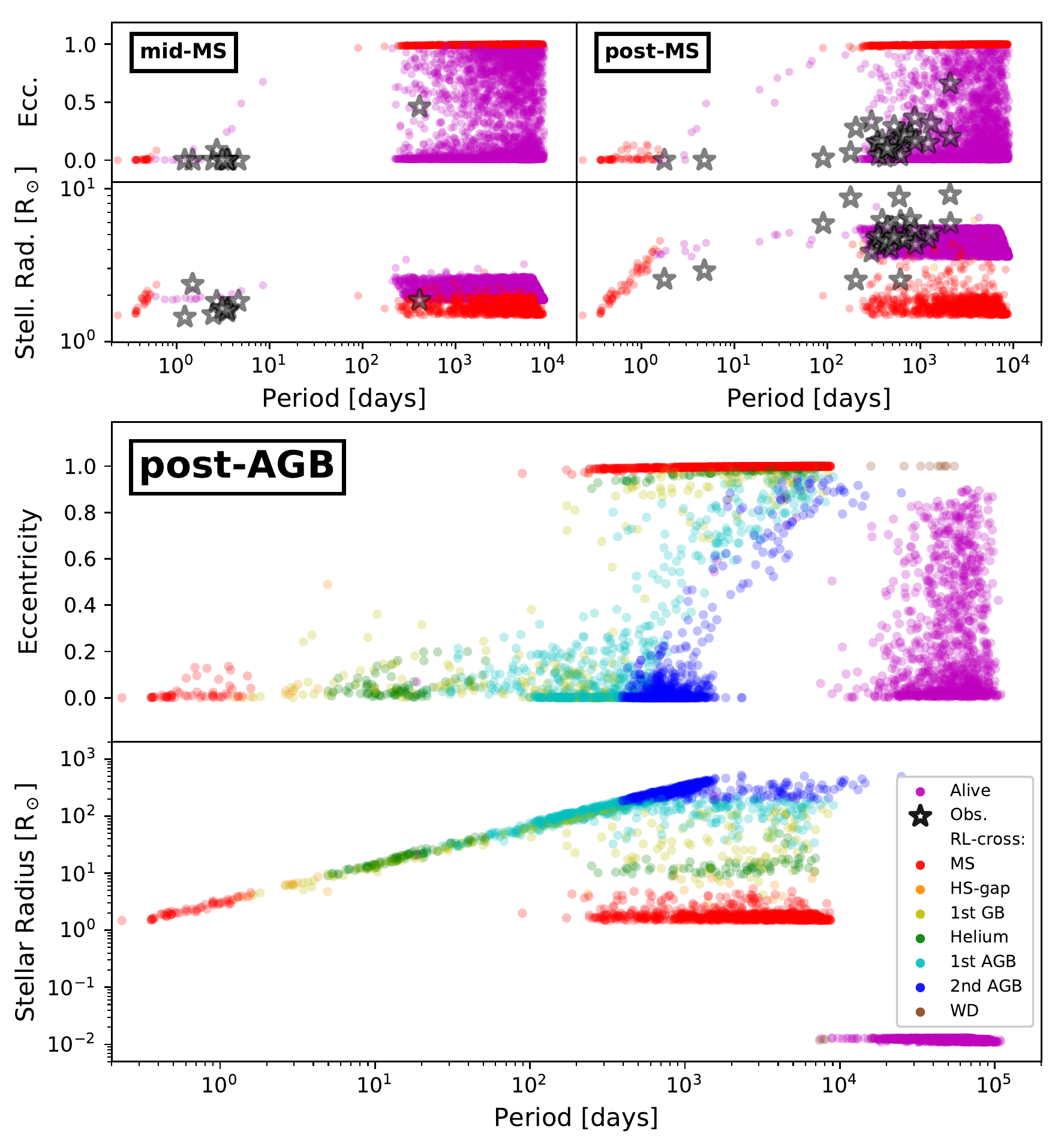}
\caption{{\bf Dynamical evolution of Jupiters around A-type stars in binary systems.} Top frame of each panel: Eccentricity vs.\ Orbital Period. Bottom frame of each panel: Stellar Radius vs.\ Orbital Period. The magenta dots show the parameters of Jupiters that survive to {either} the middle of the main sequence (labeled mid-MS), the beginning of the post-main sequence (labeled post-MS), or the white dwarf phase (labeled post-AGB). Differently colored dots show the final parameters for Jupiters that were destroyed by their stars either through EKL-induced high eccentricity Roche limit crossing or engulfment during stellar expansion. The colors represent the various stellar evolution phases at the time of planetary death: red - main sequence, orange - Hertzsprung gap, yellow - First Giant Branch, green - Core Helium burning, cyan - First Asymptotic Giant Branch, blue - Second Asymptotic Giant Branch, and brown - White Dwarf phase. Out of 4070 planets, about 2870 were destroyed before 13 Gyr had passed, falling into two distinct main groups - (1) high eccentricity, KL-driven deaths, and (2) low eccentricity, tidally and stellar expansion driven deaths. The black stars are showing the positions of some observed Jupiters around A-type stars in this parameter space, see Section \ref{sec:HJ} for references.}
   \label{fig:LargePlot}
\end{figure*}

\section{Results}\label{sec:res}
\subsection{Classification of Dynamical Evolution Outcomes}
From our $4{,}070$ simulated systems, about $70~\%$ ended in the destruction of the Jupiter-like planet, while in only $30~\%$ of cases did the planet survive to the WD phase of its host star. We identify several distinct groups of orbital evolution behaviors and outcomes for Jupiters around A-type stars in stellar binaries. There are mainly four such groups (see Table \ref{tab:percentages} for percentages): 

\begin{enumerate}
\item {\bf Classical Hot Jupiters:} These are HJs on orbits shorter than about 10 days, as one would expect from previous studies of high-eccentricity migration \citep[e.g.,][]{Dan,Wu+07,Naoz11,Naoz+12bin,Petrovich2015}. These planets reach their short period orbits during the host stars' main sequence lifetime due to an interplay of EKL-caused high orbital eccentricities and tidal effects. They are ultimately engulfed and destroyed as the stars evolve and expand. About $1.5~\%$ of our systems experience this outcome. We further discuss this group in Sec.\ \ref{sec:HJ}. 

\item {\bf Roche-limit crossers:} These are those planets that reach extremely large eccentricities through the EKL mechanism and cross their host stars' Roche limits or graze the stars' surfaces. {We assume that they are destroyed upon crossing the Roche limit and end the computation of their orbital evolution.} The actual fate of these planets might be more complicated, and some might even survive {for an extended time after crossing the Roche limit \citep[e.g.,][]{Faber+2005, Dosopoulou+17,MacLeod+18}}, however, for simplicity we mark them all as ``RL-cross'' in Fig.\ \ref{fig:LargePlot}. {Some of the possible effects are discussed in Sec.\ \ref{sec:effects}.}
About $31~\%$ of our systems lead to Roche-limit crossing, $23~\%$ of which occur during main sequence and $8~\%$ occur during post-main sequence evolution. Those that cross the stellar Roche-limit during post-main sequence evolution simply do not undergo high enough eccentricity excitations or are on initial orbits too wide to have short periapsis distances during the stellar main-sequence lifetime.

\item {\bf ``Temporary'' Hot Jupiters:} These are planets that did not reach high enough eccentricities during the main sequence lifetime of their host stars to experience tidal circularization and orbital shrinking, but which do so as the stars leave the main sequence and become giant stars. Virtually all of them only classify as ``hot'' Jupiters (in terms of temperature) for a short part of their total lifetime. They get engulfed as their host stars continue to expand and as tides continue to drive them to the stellar surface{, however their orbits usually do not fully circularize before engulfment}. The engulfment process may result in energetic disturbances on the host star and might serve as an observable \citep[e.g.,][]{MacLeod+18} {(see also Sec.\ \ref{sec:effects})}. We find that about $37~\%$ of our systems experience this outcome, however about a fifth of these ($7~\%$ of all systems) reach this outcome even though the EKL perturbations by their companion stars are negligible, as their initial orbits are relatively close to the star, at about $1-3$~AU. Further details about this group are discussed in Sec.~\ref{sec:THJ}. 

\item {\bf Surviving Jupiters:} These are planets that never significantly interact with their host stars and survive until the stars become WDs. For this population, the companion stars' EKL perturbations were too weak to cause large eccentricities, and the planets' orbits were too wide to experience strong tidal effects during post-main sequence evolution. {\it This population mostly reflects our lack of knowledge on the initial conditions of these systems.} About $30~\%$ of Jupiters survive to this stage. A small fraction of these ($0.3~\%$ of all systems) gets destroyed and accretes onto the WDs as the stellar mass loss changes the orbital parameters of the systems, allowing extreme excitations of the orbital eccentricities through the EKL mechanism, ultimately driving these Jupiters to cross their own Roche limits and to get tidally disrupted by the WDs. Examples of this potential WD pollution mechanism have been discussed in detail in \citet{Stephan+2017}. Recent work \citep{vanLieshout+2018} also indicates that some of the planet's material can be recycled into new planets and escape accretion. 

\end{enumerate}

\begin{table}
	\centering
	\caption{{\bf Jupiter evolution outcome percentages.} Listed are Classical HJ (CHJ), RL-crossing Jupiter (RL-cross), Temporary HJ (THJ), and surviving  Jupiter (Survived), up to Hubble time, outcomes as percentages of the whole population of simulated systems. The percentages are given for the whole evolution of the stars (Total), and also split between main sequence (MS), post-main sequence (Post-MS), and White Dwarf (WD) phases.}
	\label{tab:percentages}
	\begin{tabular}{ lcccc } 
		\hline
		Phase & CHJ & RL-cross & THJ & Survived \\
		&            &          &   &up to $ t_{\rm Hubble}$\\
		\hline
		Total & 1.5\% & 31\% & 37\% & 30\%\\
		\hline
		MS & 1.5\% & 23\% & - & -\\
		Post-MS & - & 8\% & 37\% & -\\
        WD & - & 0.3\% & - & 30\%\\
		\hline
	\end{tabular}
\end{table}

\begin{figure*}
\hspace{0.0\linewidth}
\includegraphics[width=1.\linewidth ]{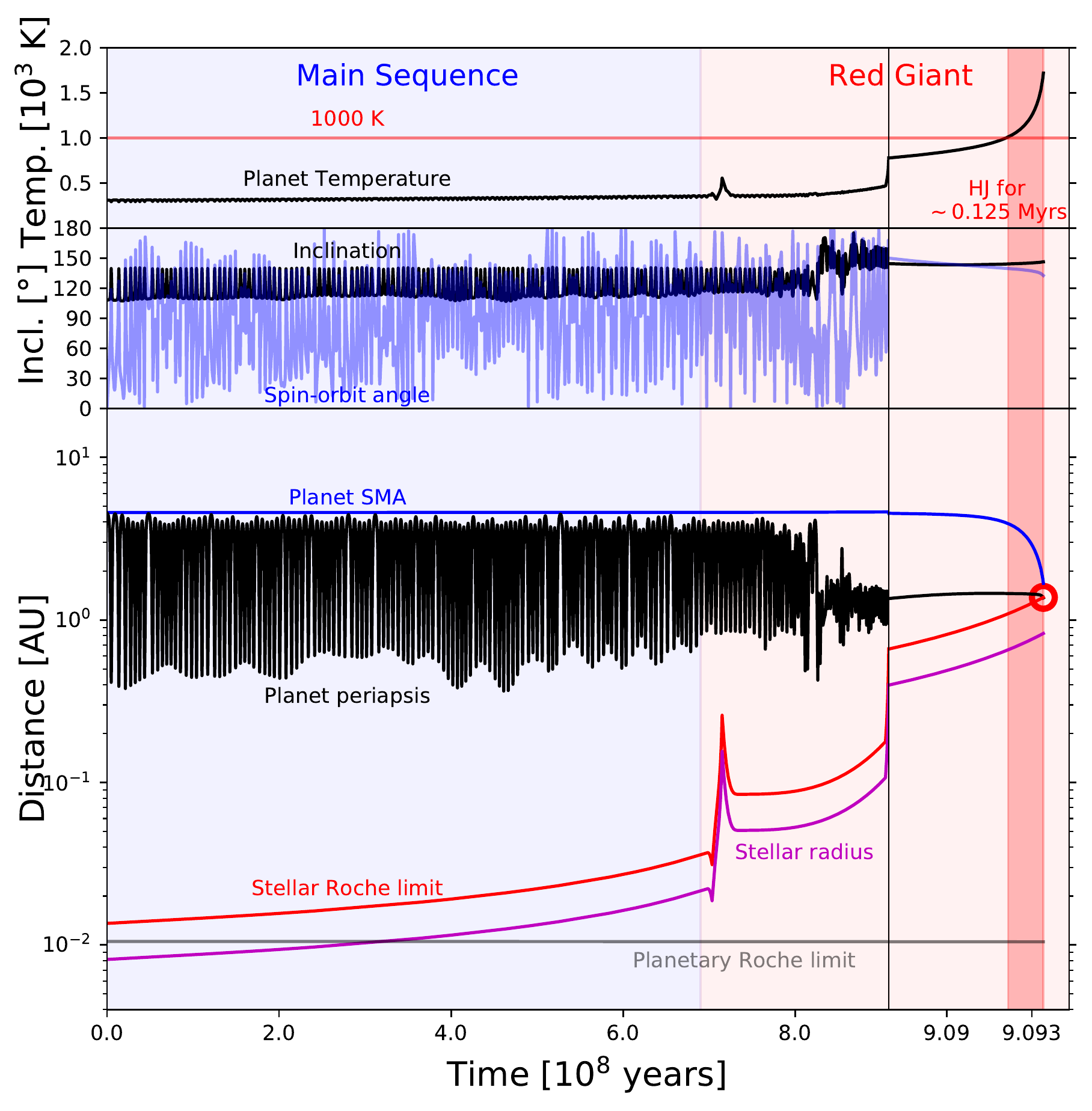}
\caption{{\bf Orbital evolution of a Jupiter around an evolving A-type star in a stellar binary, leading to formation and destruction of a Temporary Hot Jupiter.} The figure shows an example Jupiter's equilibrium temperature (top frames), inclination and spin-orbit angle (middle frames) and semi-major axis, periapsis, and stellar radius evolution (bottom frames) over time. In the bottom frames, the blue line shows the planet's semi-major axis, red and magenta show the host star's Roche limit and radius, and grey shows the planet's Roche limit. The blue shaded region marks the host star's main sequence phase, while the red shaded area marks the post-main sequence phase. The left frames of the figure show the first $908.8$~Myr of evolution, while the right frames focus on the last $0.5$~Myr before the planet enters the star's Roche lobe. Note that the planet's semi-major axis undergoes rapid tidal decay once the red giant star has expanded sufficiently, on the very right edge of the figure. This rapid orbital decay lasts on the order of $300,000$ years before the planet reaches the stellar Roche limit. The equilibrium temperature of the planet rises above $1000$~K (marked by the red line in the top frames) for the last $125,000$ years before entering the Roche lobe, rapidly increasing as the orbit decays, as highlighted by the darker red shaded area. Initial system parameters are $m_{*,1}=2.39$~$M_\odot$,  $m_{*,2}=1.95$~$M_\odot$, $a_1=4.58$~AU, $a_2=601.6$~AU, $e_1=0.01$, $e_2=0.587$, spin-orbit angle$=0\degree$, and $i=108.2\degree$.}
   \label{fig:EvoPlot}
\end{figure*}

In Fig.\ \ref{fig:LargePlot} we show three snapshots of the different systems' realizations during the stars' lifetimes, mid-main sequence (top left), at the beginning of post-main sequence (top right), and post ABG phase (bottom). We consider the eccentricity as a function of the planet's period (top panels). For comparison, we also plot the observed systems (depicted as black stars) for the relevant snapshot. For context, we also show the stellar radii for all of our systems (bottom panels). When a planet crossed the Roche-limit, we stop the integration and mark its orbital parameters. 

{Note the slight offset between the distribution of red and magenta points on the right side of the lower frame of the mid-main sequence panel. This offset is simply caused by the red points showing planets that have been destroyed via EKL-driven high-eccentricity Roche lobe crossing. This happens quickly, before the host stars can significantly evolve and expand. All the planets that lie in the region of the parameter space that allow such large eccentricities to be reached do so and get tidally destroyed or engulfed early in the systems' lifetime. By the middle of the main sequence lifetime of the star, the rest of the systems have evolved more and their stars have expanded, however there simply are not many planets left in the correct parameter space to cross the new, expanded Roche limit; they have all already been destroyed.  However, once the stars swell to become red giants, there is a new phase of destruction of planets that did not reach extreme eccentricities. The difference in behavior between these two different populations is due to the distinction between ``quadrupole'' and ``octupole'' types of EKL evolution \citep[see][for a full review of the EKL mechanism]{Naoz2016}. Thus, we predict an early ``burst'' of planet destruction during the early main sequence evolution, followed by a pause until the stars become red giants, followed by another phase of planet destruction throughout post-main sequence evolution.}


\subsection{Classical Hot Jupiters and Surviving Jupiters}\label{sec:HJ}
As we show in Fig.\ \ref{fig:LargePlot} (see upper panels), by the middle of the main sequence lifetime of the host stars, several classical HJs have been formed through high-eccentricity migration (shown in the upper left panels as purple dots, with periods shorter than about $10$~days and eccentricities close to $0$), or are still in the process of forming until the end of the main sequence. However, these more massive stars expand their radius during the main sequence by about a factor of two, which results in a higher rate of Roche-limit crossing than for less massive stars, since the more massive stars evolve and expand more quickly and the Roche-limit is proportional to the radius of the star (see equation \ref{eq:Roche}). In total, we form $64$~such HJs, or about $1.5~\%$ of our systems, which is broadly consistent with{, though somewhat less efficient than,} previous estimates of EKL-induced high-eccentricity migration models in stellar binaries{, which considered smaller mass host stars} \citep[see, for example][]{Naoz+12bin,Petrovich2015,Naoz2016,Anderson+2016}. Some of these classical HJs only survive for short times at these orbits, as the interplay of the EKL mechanism and tides keeps driving them towards the stellar surface. However, many can exist for tens to hundreds of Myr after formation. Ultimately though, all classical HJs get engulfed and destroyed as their host stars evolve and expand in radius. By the end of the main sequence, most of them have been destroyed (see upper right panels of Fig.\ \ref{fig:LargePlot}; red dots show destroyed planets, with several red dots forming a line in the stellar radius vs.\ orbital period parameter space where HJs were engulfed by their expanding host stars), with the remaining ones being destroyed as the stars evolve towards the giant phase (see the orange and yellow dots in the lower, large panels). {In general, classical HJs around A-type main sequence stars form at a lower frequency than HJs around smaller mass stars through high-eccentricity migration, and exist for shorter times due to faster stellar evolution, expansion, and tides.}

The black stars in Fig.\ \ref{fig:LargePlot} show the parameters of several observed Jupiters around A-type or retired A-type stars \citep{Johnson+2007,Johnson+2008,Cameron+2010,Johnson+2010A,Johnson+2010B,Johnson+2011A,Johnson+2011B,Sato+2013,Bieryla+2014,Wittenmyer+2014,MarshallJohnson+2014,Hartman+2015,Wittenmeyer+2015B,Wittenmeyer+2015A,Zhou+2016,Borgniet+2017,Beatty+2017,Gaudi+2017,Lund+2017,Talens+2018,MarshallJohnson+2018,Siverd+2018}, and they appear to be broadly consistent with our calculations. The observed HJs agree well with our predicted parameters at some time during the middle of the main sequence lifetime of A-type stars (upper left panels), while observed Jupiters on wider orbits (periods beyond about $100$~days) around retired A-type stars agree well with our predicted parameters for Jupiters that survive to the end of the main sequence lifetime (upper right panels). Note that both observations and our predictions do not seem to show many Jupiters on intermediate orbits between $10$~and $100$~days. HJs migrate relatively fast through this part of the parameter space, on the order of a few million years (a very small fraction of the stars' total main sequence lifetime), and are unlikely to be randomly observable.

\subsection{Temporary Hot Jupiters}\label{sec:THJ}

During post main-sequence evolution, stars begin to expand, and those planets that have short pericenter distances are engulfed by their host stars.  Furthermore, our calculations show that a large number of moderately eccentric giant planets will undergo significant tidal interactions with the expanding red giant stars, leading to orbital shrinking and circularization before the eventual crossing of the Roche lobe or engulfment by the star, as shown by the example system evolution in Fig.\ \ref{fig:EvoPlot}. {During the red giant phase, the strength of tidal effects on the star increases significantly due to the stellar radial expansion and the increased size of the convective envelope. During their main-sequence lifetime, A-type stars are mostly radiative, which severely reduces the effectiveness of their tides \citep{Zahn1977}. We consequently observe a switch in the role of tides between the main sequence and post-main sequence evolution; while during the main sequence, the tidal effects on the planet were more significant, during post-main sequence evolution, the tides on the star dominate.} Tidal interactions and planet engulfment can be expected to have significant effects on the red giant stars' envelope and mass loss evolution, as well as changing the stellar rotation rate; such giant planet interactions with evolved stars have been used to explain observed irregularities in the shape of the horizontal branch in the Hertzsprung-Russel diagram \citep[e.g.,][]{Soker1998,Soker+2000,Livio+2002}. Our ``Temporary'' HJs would most likely lead to such interactions, though we stop our calculations at the entering of the stellar Roche lobe. Further investigation would require a full hydro-dynamical treatment of the star-planet interactions. We assume that the further evolution of the stellar envelope and post-AGB remnant will be altered due to these interactions. THJ engulfment could also lead to lithium enrichment in the giant stars' atmospheres \citep[e.g.,][]{Aguilera+2016}. 

\begin{figure*}
\hspace{0.0\linewidth}
\includegraphics[width=1.\linewidth ]{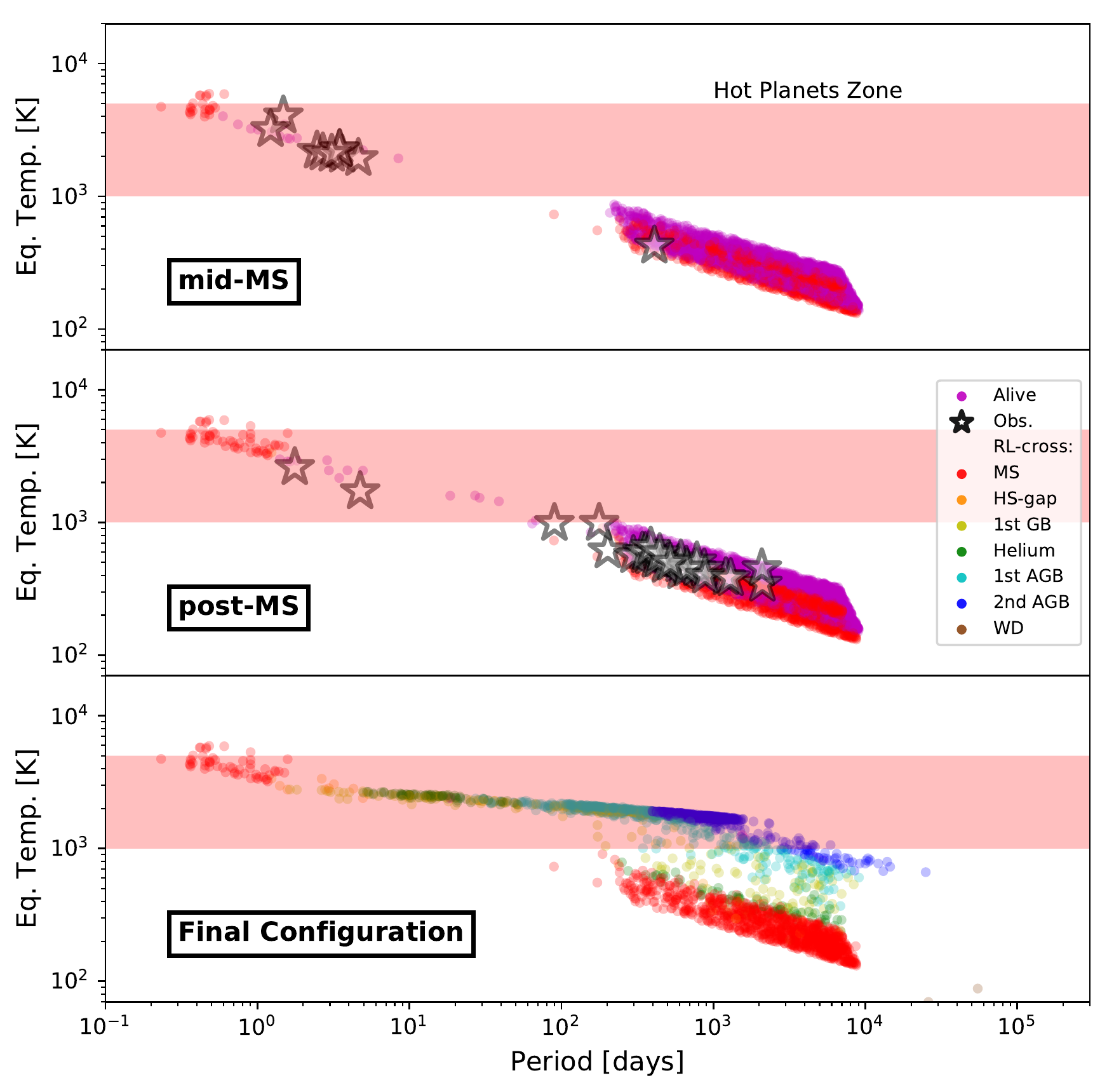}
\caption{{\bf Equilibrium temperatures of Jupiters around A-type stars in binary systems.} Equilibrium temperature vs.\ period of Jupiters for three evolutionary periods: middle of main sequence (upper panel), beginning of post-main sequence (center panel), and WD phase (bottom panel). Values for observed Jupiters around A-type stars are included as black stars. Dot colors have the same meaning as in Fig.\ \ref{fig:LargePlot}. The red shaded area shows the approximate parameter space of ``hot'' planet temperatures, from around $1{,}000$ to $5{,}000$~K. Note that THJs, which are within the red shaded area in the lower panel, can reach final equilibrium temperatures of several thousand Kelvin shortly before being destroyed, due to the intense stellar luminosity of the post-main sequence host stars. Equilibrium temperatures were calculated using equation \ref{eq:Teq} \citep{Mendez+2017} for eccentric orbits, assuming zero albedo, or taken from papers cited in section \ref{sec:HJ} for observed planets.}
   \label{fig:TempPlot}
\end{figure*}

The lower large panels in Fig.\ \ref{fig:LargePlot} show the final orbital parameters of our systems in eccentricity vs.\ period (upper frame) and host-star radius vs.\ period space (lower frame). Those dots colored red, yellow, green, cyan, and blue show the final parameters before planets enter the stellar Roche lobe during the different post-main sequence and pre-WD phases. The vast majority of these, those that form a distinct line on the left in the stellar radius vs.\ period frame, experience at least some degree of high-eccentricity migration, and reach equilibrium temperatures comparable to classical HJs (see Fig.\ \ref{fig:TempPlot}). These planets begin to experience significant tidal interactions with their host stars as those stars expand in radius, facilitated by increasing orbital eccentricities due to the EKL mechanism. However, due to the continued stellar evolution of the host stars and the significantly enhanced tides, most of these THJs enter their host stars' Roche lobe even before they can be fully circularized, making their status as HJs very short-lived. We therefore name this class of HJs ``Temporary'' Hot Jupiters (THJs). The THJ shown in Fig.\ \ref{fig:EvoPlot} only lives for about $300{,}000$ years once tidal orbital decay becomes efficient before it reaches the stellar Roche limit. This process should, as mentioned above, have significant effects on the further evolution of the giant star's outer envelope. 

\begin{figure*}
\hspace{0.0\linewidth}
\includegraphics[width=1.\linewidth ]{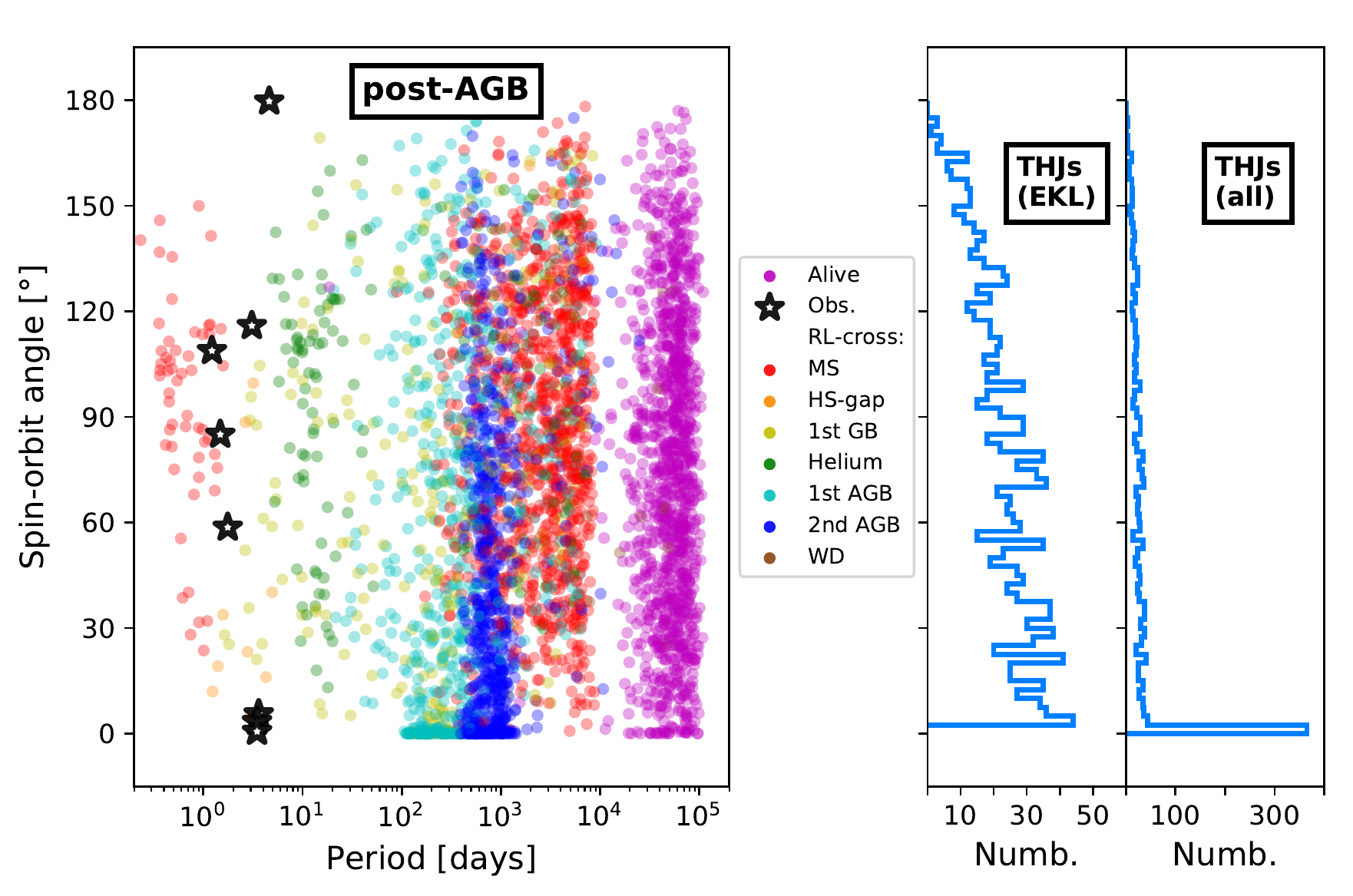}
\caption{{\bf Spin-orbit misalignment angles for Jupiters around A-type stars in binary systems.} Left large panel: Spin-orbit misalignment angle vs.\ period for Jupiters during the WD phase, with magenta dots showing survived Jupiters and differently colored dots showing destroyed Jupiters. Dot colors have the same meaning as in Fig.\ \ref{fig:LargePlot}. The plot shows that HJs, both classical and temporary, that form due to the EKL mechanism are most likely to have significantly misaligned orbits compared to their host stars' spin axes. This is broadly consistent with the few measured {\it projected} spin-orbit angles for HJs around A-type stars, shown by the black stars.  THJs that form without the influence of the EKL mechanism are, in contrast, very well aligned (depicted by the blue and cyan points that are clustered near $0\degree$). The two histograms in the right side panels show the distribution of misalignment angles for THJs. The far right histogram, labeled ``THJs (all)'', shows the spin-orbit angle distribution of all THJs, including both those formed with and without EKL effects, while the histogram to the left of it, labeled ``THJs (EKL)'', only shows those formed through EKL effects. Both histograms are practically identical except for the scale of the x-axis, which represents the number of simulated systems. The large peak at $0\degree$ inclination in the far right panel simply includes all non-EKL THJs. About $20~\%$ of THJs are in this peak and form without EKL effects, while about $80~\%$ of our THJs form through the EKL mechanism, spread across the rest of the spin-orbit angle parameter space.}
   \label{fig:ObliPlot}
\end{figure*}

\begin{figure}
\hspace{0.0\linewidth}
\includegraphics[width=\linewidth ]{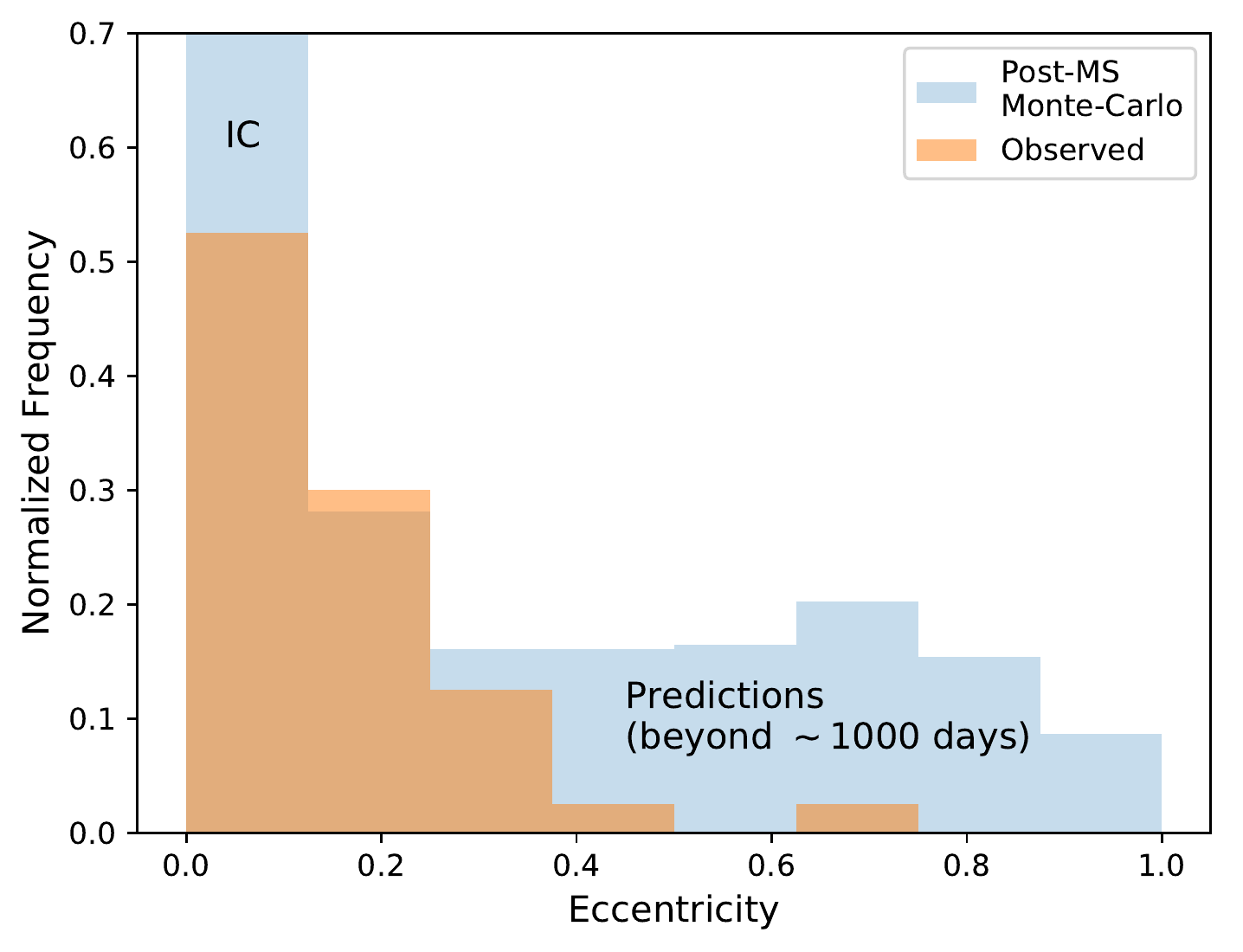}
\caption{{\bf Simulated vs.\ observed eccentricity distributions for Jupiters around retired A-type stars.} The plot shows the normalized frequencies of Jupiter eccentricities from observations (in orange) and from our simulations (in blue) during the post-main sequence phase. Note that the high peak between eccentricities of $0$ and $0.1$ in the post-MS distribution continues beyond the frame; as we assumed very small eccentricities for all Jupiters at the beginning of our simulations, this is probably an artifact of our initial conditions. Furthermore, our simulations predict a more uniform distribution of eccentricities that continues towards higher values than currently observed. However, these higher eccentricity planets exist mostly beyond periods of $\sim1000$ days, making them very difficult to observe {with current methods} (see Fig.\ \ref{fig:LargePlot}, top right frame). {Note that the currently highest eccentricity Jupiter observed around a ``retired'' A-type star does indeed have an orbital period of about 2000 days \citep{Sato+2013}.}}
   \label{fig:EccHist}
\end{figure}

While most ($\sim80~\%$) of the THJs we discuss here are products of high eccentricity migration caused by the EKL mechanism and tidal dissipation and therefore depend on the presence of a companion star, some THJs can also be formed without a companion. If a gas giant's initial orbit is sufficiently close, in general smaller than about $3$~AU, post-main sequence expansion of the host star will eventually lead to tidal interactions and engulfment of the planet even without increased orbital eccentricity, consistent with previous works investigating the evolution of gas giants around giant stars \citep[e.g.,][]{Villaver+2009, Spiegel+2012, Lopez+2016}. However, one can distinguish between these two cases considering the spin-orbit alignment of the planet and the main star. If a THJ is formed without companion excitations, the orbital plane will not change from its initial orientation, which we assume to be aligned with the stellar spin. If the THJ was formed through the EKL mechanism, however, the spin-orbit angle will be from a nearly uniformly random distribution between $0{\degree}$ and $180{\degree}$. We can see this in Fig.\ \ref{fig:ObliPlot}. The two histograms on the right side of the figure show the distribution of spin-orbit angles among THJs. Both histograms show the same population, only that the far right one labeled ``THJs (all)'' includes THJs formed both with and without EKL effects, while the one labeled ``THJs (EKL)'' only shows those formed through EKL effects. Those THJs formed without EKL effects, for which the stellar companion was irrelevant, are all aligned and form the large peak at $0^{\degree}$, while those formed with EKL effects are nearly evenly spread across the whole range of angles between $0{\degree}$ and $180{\degree}$. While the peak is large, it is extremely narrow and only includes $\sim20~\%$ of our total number of THJs.

\section{Observational Signatures}\label{sec:obs}
\subsection{Equilibrium Temperature}

The equilibrium temperature of planets is a potentially observable signature \citep[e.g.,][]{Gaudi+2017}. We calculate the time-averaged equilibrium temperature for elliptical orbits for the planets in our simulations, following equations by \citet{Mendez+2017}, namely:
\begin{equation}
    \langle T_{eq} \rangle \approx T_0  \left[ \frac{(1-A)L}{\beta \epsilon a^2} \right] ^\frac{1}{4} \left[ 1-\frac{1}{16}e^2-\frac{15}{1024}e^4 + \mathcal{O}\left(e^6\right) \right],
    \label{eq:Teq}
\end{equation}
where $T_0=278.5$~K, Earth's equilibrium temperature, $A$ is the planetary albedo (assumed to be zero), $L$ is the host star's luminosity (in $L_\odot$), $e$ and $a$ are the planet orbit's eccentricity and semi-major axis (in AU), respectively, and $\beta$ and $\epsilon$ are coefficients for the planet's heat distribution and emissivity, respectively (both assumed to be $\sim1$ for simplicity).

The CHJs in our calculations have typical temperatures of about $2{,}000$ to $5{,}000$~K, consistent with estimated temperatures of observed HJs, as can be seen in the top frame of Fig.\ \ref{fig:TempPlot}, which shows our CHJs as magenta dots and observed HJs as black stars in the ``Hot Planets Zone''. CHJs remain at these temperatures for potentially millions of years, until they either evaporate or are engulfed by the host star as it leaves the main sequence (see the red dots in the middle frame of Fig.\ \ref{fig:TempPlot}. We find that THJs reach temperatures between $1{,}000$ to $3{,}000$~K just before entering the stellar Roche lobe, as shown in the bottom frame of Fig.\ \ref{fig:TempPlot}. The THJs only exist at such high temperatures for a short time, on the order of a few $100{,}000$ years, as seen in the top right frame of Fig.\ \ref{fig:EvoPlot}, as the increase in temperature follows from the planets' rapid orbital decay and the rapid increase in stellar luminosity due to post-main sequence stellar evolution. Considering that the stellar luminosity increases tremendously during post-main sequence stellar evolution, even THJs will produce a rather small contribution to the overall stellar spectra. At a wavelength of around $1~\mu m$, the contrast between THJ and stellar emissions will be around $10^{-5}-10^{-7}$. Additionally, the expanding stars' large size will make planet transit detection signals smaller as well, however the larger size will also increase the chance that a transit can occur. At least for THJs around ``small'' giant stars ($R_*\sim5~R_\odot$), such as for stars during the helium burning phase, models suggest that transits are indeed observable \citep[e.g.,][]{Assef+2009}, and some giant planets transiting giant stars of such sizes have indeed been observed \citep[e.g.,][]{Lillo-Box+2014}, though it remains unclear if such transits would currently be observable for stars with sizes of hundreds of solar radii. {However, if the increase in temperature leads to significant planetary inflation and if stellar high energy radiation or winds can drive a significant mass loss from the Jupiter, absorption lines from the stripped planetary material might be observable \citep[e.g.,][]{RMC+2009}.}

\subsection{Stellar Obliquity} 

Detecting planets through the transit method makes it possible to observe the (projected) angle between the stellar spin axes and the planets' orbital planes through the Rossiter-McLaughlin effect \citep{GW07}. Our calculations show that THJs, like CHJs, should preferentially be misaligned to the stellar spin axis, with a nearly uniform distribution of spin-orbit angles, showing only a small preference against fully retrograde orbits, as can be seen from the THJ spin-orbit angle histograms in Fig.\ \ref{fig:ObliPlot}. Using the spin-orbit angle, it should also be possible to distinguish THJs born through EKL effects from THJs without that effect, as the later will remain aligned to their original spin-orbit angle, which can be assumed to be small, close to $0\degree$ in our case. As can be also seen from the left frame of Fig.\ \ref{fig:ObliPlot}, observed projected spin-orbit angles for HJs also broadly confirm a broad range of misalignment angles, potentially caused by EKL effects.

\subsection{Orbital Eccentricity}

The eccentricities of planetary orbits is an important factor for understanding the architectures of planetary systems, as larger eccentricities can be indicators of significant dynamical interactions, such as planet-planet scattering events or secular perturbations, while small eccentricity values are expected from disk models of planetary formation and a subsequently quiescent dynamical history. Massive evolved stars show a deficiency of short period eccentric planets \citep{Sato+2008,Bowler+2010,Johnson+2007,Johnson+2008,Johnson+2010A,Johnson+2010A,Johnson+2013,SchlaufmanWinn2013}. Our calculations show that this feature is in agreement with the dynamical evolution of giant planets in stellar binaries. As shown in Fig.\ \ref{fig:LargePlot}, top panels, both the MS-phase as well as the post-MS phase are in agreement with the observed giant planet eccentricities (the latter depicted as black stars). {We predict that there is a large population of highly eccentric planets around main sequence and retired A-type stars, as shown in Fig.\ \ref{fig:EccHist} by the blue histogram; however, most of these high eccentricity planets have orbits longer than $\sim1000$ days (see Fig.\ \ref{fig:LargePlot}, top panels), making them difficult to observe with current methods. Interestingly, the highest eccentricity {Jupiter sized} planet observed that orbits a retired A-type star, HD~120084, is very consistent with these predictions, having an eccentricity of $0.66$ and a period of $2082$~days \citep{Sato+2013}.}

{Our results also indicate that most THJs will migrate to close-in orbits and get engulfed by their expanding host stars before the orbits can fully circularize, as seen in Fig.\ \ref{fig:LargePlot}. These planets will still have some residual eccentricities up until engulfment, consistent with a recent study of close-in planets around evolved stars \citep{Grunblatt+2018}.}

 In our simulations, we assume that all giant planets start with very small eccentricities and treat them as singular planets around their host stars. This ignores the potential for planet-planet scattering during the further evolution. In the normalized histograms in Fig.\ \ref{fig:EccHist}, we see that this artifact of our initial conditions produces an extremely high peak at small eccentricities for our simulated systems during the post-main sequence phase (see blue histogram). This peak simply represents those systems that lie outside the parameter space that can undergo EKL oscillations. We propose that the actual initial eccentricity distribution should be more broad, probably caused by planet-planet interactions after dissipation of the planet-forming disk.


\subsection{Effects of Post-MS Roche Lobe Crossing and Engulfment of THJs}\label{sec:effects}

A giant planet entering an expanding post-main sequence star's Roche lobe, and subsequently the stellar envelope, can undergo and cause a multitude of effects. If the giant planet is massive and dense enough, it might resist dissipation and begin to accrete stellar material, potentially becoming a ``stellar'' companion undergoing complex common-envelope evolution with the main star \citep[e.g.,][]{Soker+1984}. Planet engulfment should also lead to the deposition of angular momentum into the star, changing the spin rate \citep[see][]{Carlberg+2009}, which has similar consequences during the main-sequence phase \citep[][]{Carlberg+2009,Qureshi+18}, as well as the deposition of energy into the stellar envelope due to drag forces and orbital decay, increasing the luminosity of the star or producing bright UV and X-ray transients \citep[e.g.,][]{Metzger+2012,MacLeod+18}. The strength of drag forces depends especially on the density of the stellar envelope and the orbital speed of the engulfed planet, which would lead to different strengths of this effect between main sequence and post-main sequence stars or between eccentric and circular planetary orbits. Furthermore, giant planets can have a range of masses beyond the simple Jupiter-analogs we have considered here, and can vary in density, especially if the planets are inflated due to increased temperatures \citep{Lopez+2016}. This would lead to further variations in the possible merger outcomes \citep[e.g.,][]{Metzger+2012,Siverd+12}, while also producing lithium enrichment through the engulfed giant planets of varying strengths \citep[e.g.,][]{Aguilera+2016}. {The engulfment process could also lead to the ejection of material from the stellar envelope or planet, leading to the formation of dust around the star. A potential candidate for this scenario is the first-ascent giant star TYC 4144 329 2, which also has a wide separation binary companion consistent with our model \citep{Melis+2009}.} Overall, our predicted THJs should have a, potentially significant, influence on the evolution of their host stars post-engulfment, which could present another observable signature of THJs through increased spin rates, larger stellar luminosities, {dust formation,} and lithium enrichment. The effects on the further stellar evolution should also be more long-lived than the predicted few $100{,}000$~years of existence as THJs, making indirect detection of THJs more promising than direct detection. 

\subsection{Occurrence Rate of HJs and THJs}

To gain a better understanding of the importance of HJs and THJs as part of the overall planet population we can estimate the fraction of systems that will produce HJs or THJs in the following way:

\begin{equation}
    f_{\rm outcome}=f_b f_p f_{\rm event},
    \label{eq:fractions}
\end{equation}

where $f_b$ is the fraction of stars in binary systems, close to $100\%$ for A-type stars \citep[e.g.,][]{Raghavan+10,Moe+2017}, $f_p$ is the fraction of Jupiter-mass planets formed at distances of a few AU from their stars, which is highly uncertain, and which we extrapolate here to be $f_p\sim 0.07-0.1$ from values for G-type stars \citep[e.g.,][]{Wright+2012,Bowler2016}. Lastly, $f_{\rm event}$  is the fraction of simulated systems that has undergone one of the possible events specified in Table \ref{tab:percentages}.  For example, the percentage of systems that form HJs (during the main sequence) of all A-type stars is $f_{HJ}\sim0.15~\%$ ($\sim10~\%$ of stars have a Jupiter, and $\sim1.5~\%$ of Jupiters become HJs), while the percentage for THJs is $f_{THJ}\sim3.7~\%$. Interestingly, $\sim2.5~\%$ of all A-type star systems will consume a Jupiter during their main-sequence lifetime and about $\sim4.5~\%$ during their post-main sequence evolution.

We estimate the number of stars in the galaxy as $N_{\star}\sim (100-400)\times10^9$, of which about $1~\%$ are in the mass range we consider here \citep[e.g.,][]{Salpeter1955}. Thus, we can also estimate the rate at which a post-main sequence Roche-limit crossing will take place in the galaxy (and might result in a luminosity or spin rate signature). To first order, considering the average lifetime of an A-type star to be on the order of $1$~Gyr and assuming roughly uniform formation and death rates, the post-MS Roche-limit crossing rate is approximately $0.045-0.18$ per year. We predict that most of these events will be caused by THJ formation and engulfment and that, given that THJs go through their orbital decay phase on the order of a few $100,000$~years, there will be a few to tens of thousands of THJs in the galaxy at any given moment. The length and strength of the increased luminosity signal in red giant stars caused by THJ engulfment and orbital energy deposition is difficult to estimate, but if it is comparable to the Kelvin-Helmholtz timescale, which is on the order of a few ten thousand years for red giants, there should be thousands of such stars with enhanced luminosities in the galaxy at any given moment. These rates indicate that THJs and their effects on post-main sequence stars should be observable and have a strong effect on the luminosity function of intermediate mass red giant stars. 

\section{Discussion}\label{sec:dis}

In this work, we have explored the dynamical evolution of single giant planets around A-type stars in hierarchical binaries. Considering initially circular planetary orbits between $1$ and $10$~AU, we identify four principal evolution outcomes:

\begin{itemize}
  \item {\bf Classical Hot Jupiters (CHJs):} Giant planets that undergo high-eccentricity migration to short period orbits ($P<10$~days) during the main sequence lifetime of the main star, caused by an interplay of the EKL mechanism and tidal effects. These planets can typically reach temperatures of $2{,}000-5{,}000$~K and are eventually engulfed by the star as it expands during post-main sequence stellar evolution. About $1.5~\%$ of our giant planets in binaries lead to this result.
  \item {\bf Temporary Hot Jupiters (THJs):} THJs form during post-main sequence evolution, as the stars expand. These giant planets can either form like CHJs through high eccentricity migration caused by EKL effects and tides, or their initial orbits were close enough to their stars to be eventually heated up and engulfed by their stars even with low eccentricities ($a_1\lesssim3$~AU). These planets only exist as HJs for a few $100{,}000$~years before Roche-lobe crossing and engulfment, but can have significant effects on the stellar envelope and can reach temperatures of $2{,}000-3{,}000$~K before entering the stellar Roche lobe of the expanding star. About $37~\%$ of our systems lead to this outcome.
  \item {\bf Roche-limit crossers:} These are giant planets that undergo very strong EKL effects that are too strong to be counteracted by tidal forces, thus crossing the Roche limit or grazing the stellar surface at high eccentricities and velocities. About $23~\%$ of our giant planets experience this result during the stellar main sequence, while a further $8~\%$ do so in the post-main sequence. During the AGB-phase, the stars lose a significant part of their mass, changing the orbital parameters of some systems enough to increase EKL strength significantly. This leads to about $0.3~\%$ of our giant planets to accrete and pollute the White Dwarf remnants through high eccentricity Roche-limit crossing.
  \item {\bf Surviving Jupiters:} Gas giant planets that originally had large orbital periods and did never undergo strong enough EKL effects to lead to significant interactions with their host stars. This is the case for about $30~\%$ of our systems. 
\end{itemize}

Overall, only $30\%$ of the planets will survive to the White Dwarf phase without stellar interactions, while $70\%$ will be engulfed at some point in their evolution. The engulfed planets can have significant effects on the stellar rotation rates and luminosities. The EKL mechanism greatly enhances the fraction of planets that end up being engulfed; about $80\%$ of engulfed planets have undergone significant EKL effects. Overall, we predict a THJ engulfment rate of $\sim0.045-0.18$ per year, which, depending on the length and strength of the engulfment effects onto the red giant stars' envelopes, could translate to thousands or tens of thousands of red giants with THJ engulfment effects at any given moment in our galaxy.

{From our calculation results, we also predict that there is a large population of high eccentricity giant planets around A-type stars with orbital periods $\gtrsim1000$~days, which is difficult to observe, but is consistent with the known high eccentricity giant planets.} Our results are also consistent with the observed large, nearly isotropic spread of spin-orbit misalignment angles, further suggesting that stellar binary dynamics are crucial for the understanding of giant planet orbits around A-type star.

\section*{Acknowledgements}

We thank the anonymous referee for helpful comments and questions that helped us to improve this paper. SN acknowledges the partial support from the Sloan foundation. SN and APS also acknowledge the partial support from the NSF through grant No.\ AST-1739160. We thank John Johnson for many inspiring conversations between him and SN during the time she was a postdoc at the ITC. We thank Ben Zuckerman, Michael Fitzgerald, and Ruth Murray-Clay for comments and discussions. We also thank Ahmed Qureshi for assisting in the continuing improvement of modeling the stellar spins. Calculations for this project were performed on the UCLA cluster {\it Hoffman2}.

\software{SSE \citep{Hurley+00}, Matplotlib \citep{Hunter2007}}







\appendix
\section{A-type Stars: Definitions and Evolution}\label{App:AppendixA}

\begin{figure}
\hspace{0.0\linewidth}
\includegraphics[width=1.\linewidth ]{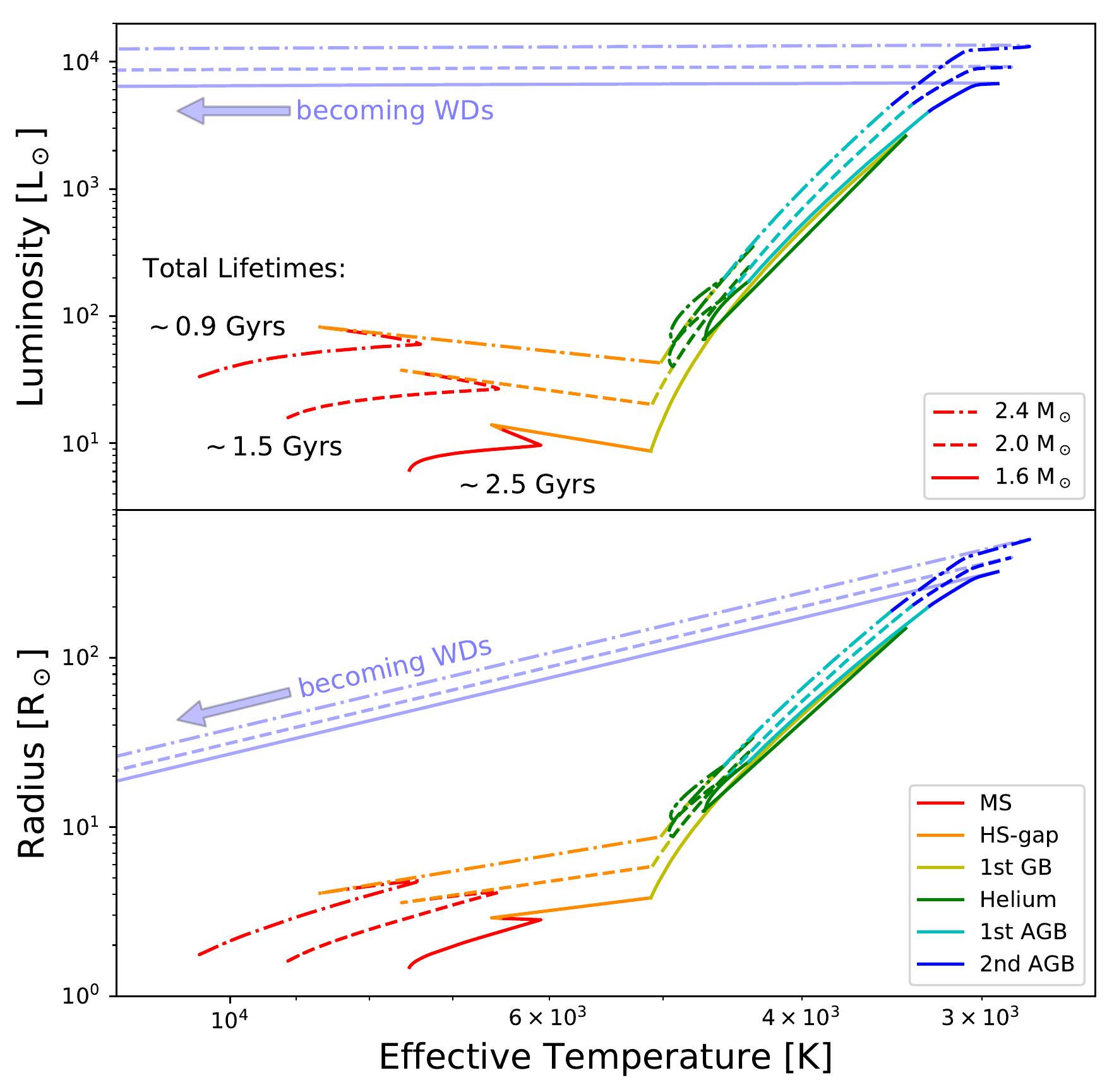}
\caption{{\bf Stellar evolution temperature vs.\ luminosity and radius profiles of A-type stars.} Shown are stellar evolutionary tracks for $1.6$, $2.0$, and $2.4$~$M_\odot$ A-type stars, shown by a line, a dashed line, and a dot-dashed line, respectively. The upper panel shows the evolution of temperature vs.\ luminosity, the lower panel shows temperature vs.\ radius. The colors of the curve segments represent different evolutionary phases, same as in Figure\ref{fig:LargePlot}: red - main sequence; orange - Hertzsprung gap; yellow - First Giant Branch; green - Core Helium burning; cyan - First Asymptotic Giant Branch; blue - Second Asymptotic Giant Branch. The light blue segment shows the evolution when the stars are becoming WDs. Lifetimes until becoming WDs for the three masses of stars are indicated in the upper panel. The lifetimes are, from low to high mass: $2.5$, $1.5$, and $0.9$~Gyr. Tracks and times were calculated using {\tt SSE} \citep{Hurley+00}.}
   \label{fig:HRPlot}
\end{figure}

Here we give a short overview of A-type stars' evolutionary phases. A-type main sequence stars are usually defined to have masses between $\sim1.6$ and $\sim2.4$~$M_\odot$ with surface temperatures between about $7{,}000$ and $10{,}000$~K \citep{Adelman2004}. Fig.\ \ref{fig:HRPlot} shows the evolution of temperature vs.\ luminosity and radius for three example star masses ($1.6$, $2.0$, and $2.4$~$M_\odot$), calculated using {\tt SSE} \citep{Hurley+00}.

As A-type stars evolve along the main sequence, shown in red in Fig.\ \ref{fig:HRPlot}, they slowly expand in radius by about a factor of two, cool down by $\sim2{,}000-3{,}000$~K, and increase in luminosity by about a factor of two as well. Note that this evolution is slightly different from sun-like G-type stars, which initially heat up during their main sequence evolution before finally cooling down. The main sequence phase lasts for about $2.2$~Gyr for $1.6$~$M_\odot$ A-type stars and $0.7$~Gyr for $2.4$~$M_\odot$. After they have expended their core hydrogen fuel, the stars then evolve through the Hertzsprung gap, shown in orange, rapidly expanding by another factor of two, cooling down to about $5{,}000$~K, and halving their luminosity over the course of $5-50$~Myr (high to low mass).

Afterwards, the First Giant Branch phase begins, shown in yellow, lasting for $6-100$~Myr. This phase progresses very differently for low vs.\ high mass A-type stars. A $1.6$~$M_\odot$ star grows from $4$ to $140$~$R_\odot$, increases in luminosity from $8$ to $2500$~$L_\odot$ and cools down from $5{,}000$ to $3{,}500$~K, while a $2.4$~$M_\odot$ star grows only from $8$ to $33$~$R_\odot$, increases in luminosity from $42$ to $350$~$L_\odot$ and cools down from $5{,}000$ to $4{,}500$~K. At the end of the First Giant Branch phase, the stars contract to their previous radius, luminosity, and temperature, as they begin to burn helium in their cores, shown in green. The helium burning phase lasts for about $200$~Myr for $1.6$~$M_\odot$ stars, $300$~Myr for $2.0$~$M_\odot$ stars, and $130$~Myr for $2.4$~$M_\odot$ stars. After expending their helium fuel, the stars rapidly evolve to become AGB giants, shown as in cyan (first ascent) and dark blue (second ascent), expanding to sizes of about $3{,}000-4{,}000$~$R_\odot$, cooling down to $3{,}000$~K, and increasing their luminosities to about $10{,}000$~$L_\odot$ over the course of $5-9$~Myr. The stars then expel their outer layers and lose mass, becoming white dwarfs.

In total, $1.6$~$M_\odot$ stars need $2.5$~Gyr, $2.0$~$M_\odot$ stars need $1.5$~Gyr, and $2.4$~$M_\odot$ stars need $0.9$~Gyr to reach the white dwarf phase.

\end{document}